\documentclass[preprint,11pt]{elsarticle}

\usepackage{amsmath,amssymb,amsthm,amsfonts}
\usepackage{mathabx}
\usepackage{mathrsfs}
\usepackage{bm}
\usepackage{graphicx}
\usepackage{float}
\usepackage{subfig}
\usepackage{color,xcolor}
\usepackage{tabularx}
\biboptions{numbers,sort&compress}
\usepackage{epsfig,psfrag}
\usepackage{tikz}
\usepackage{graphicx,psfrag}
\usepackage{hyperref}

\usepackage{changes}
\makeatletter
\@namedef{Changes@AuthorColor}{blue}
\colorlet{Changes@Color}{blue}
\makeatother
\setcitestyle{authoryear,round}

\makeatletter
\hypersetup{colorlinks=true}
\AtBeginDocument{\@ifpackageloaded{hyperref}
	{\def\@linkcolor{magenta}
		\def\@anchorcolor{black}
		\def\@citecolor{teal}
		\def\@filecolor{cyan}
		\def\@urlcolor{magenta}
		\def\@menucolor{red}
		\def\@pagecolor{cyan}
		\begingroup
		\@makeother\`%
		\@makeother\=%
		\edef\x{%
			\edef\noexpand\x{%
				\endgroup
				\noexpand\toks@{%
					\catcode 96=\noexpand\the\catcode`\noexpand\`\relax
					\catcode 61=\noexpand\the\catcode`\noexpand\=\relax
				}%
			}%
			\noexpand\x
		}%
		\x
		\@makeother\`
		\@makeother\=
	}{}}
\makeatother

\makeatletter \oddsidemargin 0.0cm \evensidemargin 0.0cm
\newcommand{\be}{\begin{equation}}
\newcommand{\en}{\end{equation}}

\def\bm#1{\mbox{\boldmath{$#1$}}}

\numberwithin{equation}{section}
\theoremstyle{plain}

\newtheorem{theorem*}{Theorem}

\theoremstyle{definition}

\newcommand{\ri}{\mathrm{i}}
\DeclareMathOperator{\tr}{tr}

\DeclareMathOperator{\sech}{sech}

\journal{Mechanics of materials}

\begin{document}
	
	\begin{frontmatter}
		
		\title{{\bf  Inhomogeneous thinning of dielectric membranes under uniaxial tension and electric fields}}
		
		\author[mymainaddress]{Xiang Yu\corref{mycorrespondingauthor}}
		\cortext[mycorrespondingauthor]{Corresponding author}
		\ead{yuxiang@dgut.edu.cn}
		\author[mysecondaryaddress]{Yibin Fu}

		\address[mymainaddress]{Department of Mathematics, School of Computer Science and Technology, Dongguan University of Technology, Dongguan, 523808, China}
		
		\address[mysecondaryaddress]{School of Computer Science and Mathematics, Keele University, Staffs ST5 5BG, UK}
		
		\begin{abstract}
	Dielectric elastomers exhibit rich electromechanical instabilities arising from the coupling between mechanical deformations and electric fields. A widely used approach for analyzing instabilities in dielectric elastomers is the Hessian stability criterion proposed by \cite{zhao2007method}, which identifies the onset of instability of a homogeneous deformation but does not determine how the deformation develops beyond the instability threshold. To address this problem, we investigate dielectric membranes subjected to uniaxial tension and an electric field. Starting from a three-dimensional nonlinear electroelastic formulation, we derive asymptotically consistent reduced models, including a membrane model and a  plate model, using the variational–asymptotic method. A linear bifurcation analysis first shows that the Hessian stability criterion is equivalent to a zero-wavenumber bifurcation condition, thereby establishing a direct connection between energy-based stability analysis and bifurcation theory. A subsequent weakly nonlinear analysis demonstrates that the zero-wavenumber bifurcation gives rise to localized necking, manifested as inhomogeneous thinning of the membrane. The weakly nonlinear predictions are further validated by numerical simulations. Furthermore, for the plane-stress configuration considered here, the membrane model accurately captures both the onset of instability and the associated localization behavior, while bending effects remain small. These results identify the deformation associated with the Hessian instability and provide a framework for understanding electromechanical instabilities in dielectric membranes.
		\end{abstract}  
		
		\begin{keyword}
		Dielectric membranes\sep Electromechanical instability\sep Hessian stability criterion \sep Inhomogeneous thinning \sep Weakly nonlinear analysis
		\end{keyword}
	\end{frontmatter}

	\section{Introduction}

Dielectric elastomers are a class of soft electroactive materials that can undergo large, reversible deformations in response to electric stimuli. Owing to their high energy density, fast response, and mechanical compliance, they have been widely explored in applications such as soft actuators \citep{pelrine2000high,carpi2010stretching,sommer2004materials,henann2013modeling}, artificial muscles \citep{brochu2010advances,yuen2025electrostatic,wang2022dielectric}, adaptive optics \citep{carpi2011bioinspired,zhang2023biomimetic}, and energy harvesting devices \citep{mckay2011soft,gu2024comprehensive,hanuhov2024energy}. Their behavior is governed by a strong coupling between mechanical deformations and electric fields, which gives rise to rich nonlinear responses \citep{suo2008nonlinear,dorfmann2005nonlinear,plante2006large,li2013giant} and a variety of electromechanical instabilities, such as pull-in instability \citep{zhao2007method,zhao2009electromechanical,zhao2014harnessing}, inhomogeneous thinning  \citep{zhou2008propagation,fu2023axisymmetric,fu2018localized,yu2025analysis,zurlo2017catastrophic,yang2022inhomogeneous}, wrinkling \citep{plante2006large,greaney2019out,de2010pull}, and creasing \citep{wang2011creasing,landis2022formation}. Understanding the onset and evolution of these instabilities is therefore a central problem in the mechanics of dielectric membranes.

A widely used approach for identifying the onset of instability is the Hessian stability criterion \citep{zhao2007method,zhao2007electromechanical}. By examining the convexity of an appropriate free energy function, this criterion provides a simple and effective condition for determining the critical state at which a homogeneous deformation loses stability.  Owing to its generality and ease of implementation, as well as the fact that it only requires the homogeneous solution, this approach has become a standard tool in the stability analysis of dielectric elastomers \citep{huang2012electromechanical,dorfmann2019instabilities,su2019tuning,su2021bending}.

In parallel, incremental (small-on-large) theories have been widely used to investigate electromechanical instabilities \citep{dorfmann2005nonlinear,dorfmann2014instabilities,bertoldi2011instabilities,fu2018reduced,broderick2020stability,su2020pattern,bahreman2022structural,goshkoderia2017electromechanical}. In this framework, infinitesimal perturbations are superposed on a finitely deformed state, leading to linearized governing equations for stability analysis. This approach is well suited for capturing spatially varying instability modes and provides detailed information on bifurcation behavior, including the wavelength and mode shape of the emerging deformation. It therefore provides information that is inaccessible to energy-based criteria such as the Hessian approach.

Despite its success, the Hessian stability criterion relies only on the homogeneous deformation and does not account for spatially varying perturbations. As a result, it identifies only the onset of instability of the homogeneous solution, without determining how the deformation develops beyond the instability threshold. It is often implicitly assumed that the subsequent deformation remains homogeneous, resulting in a rapid and uniform thinning of the membrane. However, whether the deformation instead evolves into a spatially non-uniform state remains unclear. This limitation raises an important question: what form of deformation emerges when the Hessian stability criterion is violated in dielectric membranes?

To address this question, we study a dielectric membrane subjected to uniaxial tension and an electric field, following \cite{huang2012electromechanical}. Under the plane-stress condition, the deformation in the width direction is not prescribed a priori but forms part of the solution, making the problem intrinsically three-dimensional. This significantly complicates the analytical treatment of the governing equations. To overcome this difficulty, we begin with a three-dimensional nonlinear electroelastic formulation and derive asymptotically consistent reduced models using the variational-asymptotic method \citep{berdichevskii1979variational,berdichevskii1981energy,yu2012variational,hodges2006nonlinear,yu2024asymptotically}. The resulting hierarchy of two-dimensional models, consisting of a membrane model and its higher-order plate extension,  provides a tractable framework for bifurcation and weakly nonlinear analyses.

Based on the reduced formulation, we establish a direct connection between the Hessian stability criterion and bifurcation theory. Specifically, we show that the Hessian criterion coincides exactly with a zero-wavenumber bifurcation. A subsequent weakly nonlinear analysis reveals that the bifurcation is subcritical and gives rise to localized necking deformations. The theoretical predictions are further supported by numerical simulations.  These results provide a bifurcation-based interpretation of Hessian instability in dielectric membranes. Furthermore, the membrane model is found to accurately capture both the onset of instability and the associated localization behavior. This demonstrates that, unlike in plane-strain \citep{fu2018localized} and axisymmetric configurations \citep{yu2025analysis}, localization in the present plane-stress setting can be captured without invoking bending effects.

The remainder of the paper is organized as follows. Section \ref{sec:3D} formulates the three-dimensional electroelastic problem. Section \ref{sec:red} derives a hierarchy of asymptotically consistent reduced models. Section \ref{sec:hom} analyzes homogeneous deformations and reviews the Hessian stability criterion. Section \ref{sec:lin} establishes the connection between the Hessian criterion and bifurcation theory through a linear stability analysis, and Section \ref{sec:weak} presents the corresponding weakly nonlinear analysis. Numerical validation is provided in Section \ref{sec:num}. Concluding remarks are given in Section \ref{sec:con}.

\section{Three-dimensional electroelastic formulation}\label{sec:3D}

\begin{figure}[h!]
	\centering
	\includegraphics[width=0.98\linewidth]{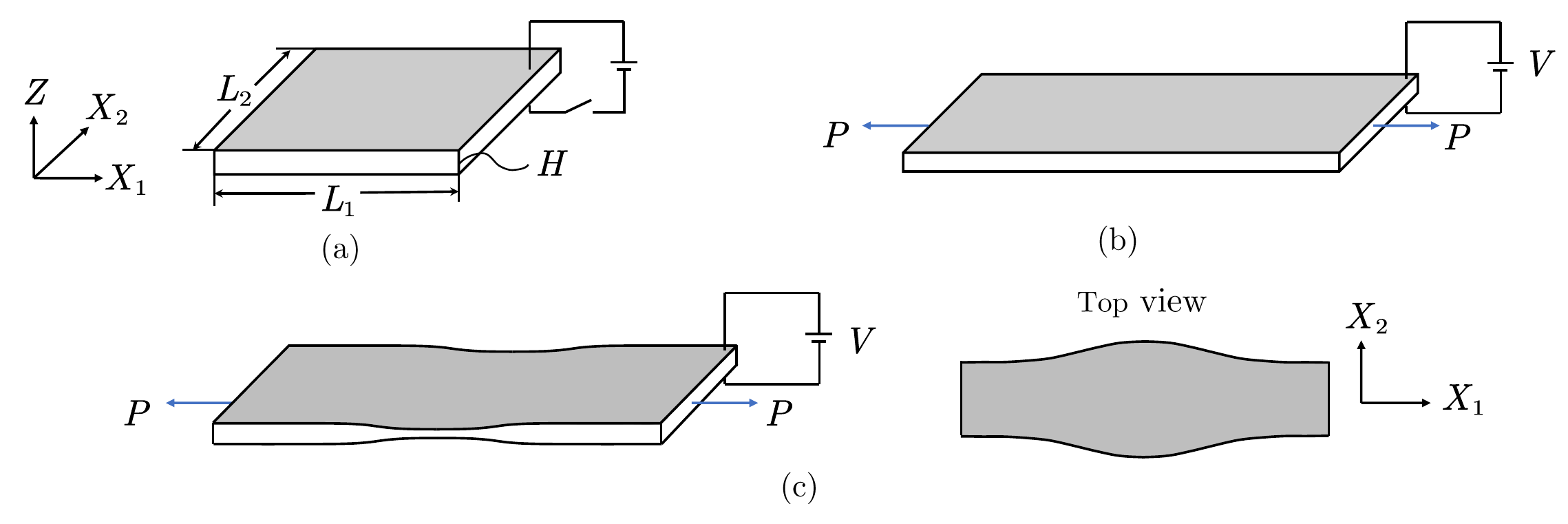}
	\caption{Three configurations of a dielectric membrane subjected to uniaxial tension and an electric field: (a) reference (undeformed) configuration, (b) homogeneously deformed configuration, and (c) current configuration together with its top view.}
	\label{fig:membrane}
\end{figure}

Consider a dielectric membrane occupying a rectangular domain of length \(L_1\), width \(L_2\), and thickness \(H\) in its reference configuration. The membrane is assumed to be thin, with \(H\) much smaller than its in-plane dimensions. It is subjected to uniaxial tension and an electric voltage is applied across its upper and lower surfaces. Depending on the loading conditions, the resulting deformation may be either homogeneous or inhomogeneous, as illustrated in Fig.~\ref{fig:membrane}.

We introduce Cartesian coordinates $(X_1,X_2,Z)$ and $(x_1,x_2,z)$ in the reference and current configurations, respectively. A material point with reference position $\bm{X}=X_\alpha \bm{e}_\alpha+Z\bm{k}$ ($\alpha=1,2$) is mapped  to the current position  $\bm{x}=x_\alpha\bm{e}_\alpha+z\bm{k}$ according to
\begin{align}
x_1=x_1(X_1,X_2,Z),\qquad x_2=x_2(X_1,X_2,Z),\qquad z=z(X_1,X_2,Z),
\end{align}
where $(\bm{e}_1,\bm{e}_2,\bm{k})$ denotes the standard basis  for the Cartesian coordinate system. The deformation  gradient is calculated as
\begin{align}\label{eq:FF}
\bm{F}=\frac{\partial\bm{x}}{\partial\bm{X}}=\frac{\partial\bm{x}}{\partial X_\alpha}\otimes\bm{e}_\alpha+\frac{\partial \bm{x}}{\partial Z}\otimes\bm{k}=\nabla\bm{x}+\bm{x}_{,Z}\otimes\bm{k},
\end{align}
where $\nabla=\bm{e}_\alpha\frac{\partial}{\partial X_\alpha}$ denotes the in-plane gradient operator and the comma signifies  partial differentiation. Under electrostatic conditions, the nominal electric field can be expressed in terms of an electrostatic potential \(\varPhi\) as
\begin{align}\label{eq:EE}
\bm{E}=-\frac{\partial\varPhi}{\partial\bm{X}}
=-\nabla\varPhi-\varPhi_{,Z}\bm{k}.
\end{align}

Following the framework of nonlinear electroelasticity of \cite{dorfmann2005nonlinear,dorfmann2014nonlinear}, the electromechanical response of the dielectric material is  characterized by a total free energy density function $\Omega(\bm{F},\bm{E})$. We further assume that the  material is incompressible, so that the deformation satisfies the constraint
\begin{align}\label{eq:incom}
\det(\bm{F})=1.
\end{align}
The associated first Piola-Kirchhoff stress tensor and nominal electric displacement are given by
\begin{align}
\bm{P}=\frac{\partial\Omega}{\partial\bm{F}}-p\bm{F}^{-T},\qquad \bm{D}=-\frac{\partial\Omega}{\partial\bm{E}},
\end{align}
where $p$ is a Lagrange multiplier enforcing the incompressibility constraint.

We adopt a variational formulation based on the total potential energy:
\begin{align}\label{eq:TE}
\mathcal{E}[\bm{x},\varPhi]=\iint_\mathcal{R}\int_{-H/2}^{H/2}\Big(\Omega(\bm{F},\bm{E})-p(\det (\bm{F})-1)-P\frac{\partial x_1}{\partial X_1}\Big)\,dZdA,
\end{align}
where $\mathcal{R}=[-L_1/2,L_1/2]\times[-L_2/2,L_2/2]$ denotes  the in-plane domain with $dA=dX_1dX_2$ and $P$ represents the applied nominal traction on the lateral surfaces in the $X_1$-direction. The electric potential is prescribed on the upper and lower surfaces  as
\begin{align}\label{eq:bc}
\varPhi|_{Z=-\frac{H}{2}}=-\frac{V}{2},\qquad \varPhi|_{Z=\frac{H}{2}}=\frac{V}{2},
\end{align}
where $V$ is the applied voltage. This three-dimensional variational formulation serves as the starting point for the subsequent analyses.

The variational principle implies the natural boundary condition $
\bm P\bm e_2=0$
on the lateral surfaces \(X_2=\pm L_2/2\). Consequently, the transverse stretch is not known a priori and must be determined as part of the solution, so that the problem remains intrinsically three-dimensional. By contrast, the plane-strain formulation considered in \cite{fu2018localized} prescribes the deformation in the \(X_2\)-direction and is effectively two-dimensional.

\section{Reduced models for dielectric membranes}\label{sec:red}

Although the three-dimensional formulation provides a complete description of the problem, its complexity renders analytical treatment difficult.   To facilitate the analysis, we derive a hierarchy of asymptotically consistent reduced models, consisting of a leading-order membrane model and a higher-order plate model that accounts for finite-thickness effects.

\subsection{Membrane model}

As a first step, we derive a membrane model that describes the in-plane electromechanical deformation.  Let $\bm R=(X_1,X_2)$ denote the in-plane reference coordinates, and define the actual middle surface by
\begin{align}\label{eq:rR}
\bm r(\bm R)
=\frac{1}{H}
\int_{-H/2}^{H/2}
\bm x(\bm R,Z)\,dZ,
\end{align}
which represents the average position through the thickness.

We assume that the middle plane is a plane of symmetry, so that the extensional and flexural modes are decoupled  \citep{fu2018reduced,su2018wrinkles}. In this work, we focus on extensional modes, for which the in-plane displacement is an even function of $Z$, whereas the out-of-plane displacement and electric potential are odd functions of $Z$. Accordingly, the deformation and electric potential can be approximated by
\begin{align}\label{eq:kinematics}
\bm{x}(\bm{R},Z)=\bm{r}(\bm{R})+a(\bm{R})Z\bm{k}+O(H^2),
\qquad
\varPhi(\bm{R},Z)=\bar{V}Z+O(H^2),
\end{align}
where $\bm{r}$ is an in-plane vector field, $a$ is the thickness stretch, and $\bar V=V/H$ is the voltage normalized by the thickness.

In view of \eqref{eq:FF} and \eqref{eq:EE}, at leading order, the deformation gradient and electric field are given by
\begin{align}\label{eq:FandE}
\bm{F}=\nabla\bm{r}+a\bm{k}\otimes\bm{k},
\qquad
\bm{E}=-\bar V\bm{k}.
\end{align} 
The incompressibility constraint \eqref{eq:incom} yields
\begin{align}\label{eq:asol}
a=\frac{1}{\det(\nabla\bm{r})},
\end{align}
where $\det(\nabla\bm r)$ is computed from the $2\times2$ matrix representation of $\nabla\bm r$. The Lagrange multiplier \(p\) associated with incompressibility is determined from the traction-free condition \(\bm P\bm k=\bm0\) on the upper and lower surfaces. Since \(\bm P\bm k\) has only a thickness component for the deformation \eqref{eq:FandE}, this condition reduces to
\begin{align}\label{eq:p0}
[\Omega_{\bm F}(\nabla\bm r+a\bm k\otimes\bm k,-\bar V\bm k)
-p(\nabla\bm r+a\bm k\otimes\bm k)^{-T}]
\bm k\cdot\bm k=0,
\end{align}
where \(\Omega_{\bm F}=\partial\Omega/\partial\bm F\). Eq. \eqref{eq:p0} determines \(p\) as a function of the in-plane deformation.

Substituting \eqref{eq:FandE} into the total potential energy \eqref{eq:TE} and normalizing by the thickness $H$,  we obtain the membrane energy in the form
\begin{align}\label{eq:mem}
\mathcal{E}_{\rm mem}[\bm r]=
\iint_{\mathcal R}
\Big(
G(\nabla\bm{r})
-P\bm e_1\cdot\frac{\partial \bm r}{\partial X_1}
\Big)
\,dA,
\end{align}
where  the effective energy density is defined by
\begin{align}
G(\nabla\bm{r}):=\Omega(\nabla\bm{r}+\det(\nabla\bm{r})^{-1}{\bm{k}\otimes\bm{k}},-\bar{V}\bm{k}).
\end{align}
 
Writing $\bm{r}(\bm{R})=(r_1(X_1,X_2),r_2(X_1,X_2),0)$, the membrane energy \eqref{eq:mem} can be expressed in component form as
\begin{align}\label{eq:memcomp}
\mathcal E_{\rm mem}[\bm r]=
\iint_{\mathcal R}
\Big(
G(r_{\alpha,X_1},r_{\alpha,X_2})
-Pr_{1,X_1}
\Big)\,dA,
\end{align}
where $\alpha=1,2$ and the comma denotes partial differentiation, e.g., $r_{\alpha,X_1}
={\partial r_\alpha}/{\partial X_1}$. Applying a variation to \eqref{eq:memcomp} yields the Euler–Lagrange equations
\begin{align}\label{eq:equi}
\frac{\partial }{\partial X_1}\Big(\frac{\partial G}{\partial r_{\alpha,X_1}}\Big)+\frac{\partial }{\partial X_2}\Big(\frac{\partial G}{\partial r_{\alpha,X_2}}\Big)=0,\qquad \alpha=1,2,
\end{align}
together with the natural boundary conditions
\begin{align}
&\frac{\partial G}{\partial r_{\alpha, X_1}}=P\delta_{\alpha 1},\qquad  X_1=\pm\frac{L_1}{2},\label{eq:bcc1}\\
&\frac{\partial G}{\partial r_{\alpha,X_2}}=0,\qquad  X_2=\pm\frac{L_2}{2},\label{eq:bcc2}
\end{align}
where $\delta_{\alpha1}$ denotes the Kronecker delta.

The membrane model is asymptotically consistent with the three-dimensional formulation as $H\to0$ and  serves as the primary reduced model in this work.

\subsection{Plate model}

To account for  bending effects, we derive a   plate model by retaining higher-order terms in the thickness expansion.

Extending the membrane approximation \eqref{eq:kinematics} to higher order in \(Z\), we write
\begin{align}
\begin{split}\label{eq:x}
&\bm{x}(\bm{R},Z)=\bm{r}(\bm{R})+a(\bm{R})Z \bm{k}+\frac{1}{2}\bm{b}(\bm{R})\Big(Z^2-\frac{H^2}{12}\Big)+\frac{1}{6}c(\bm{R}) Z^3\bm{k}+O(H^4),\\
&p(\bm{R},Z)=p_0(\bm{R})+\frac{1}{2}p_2(\bm{R})Z^2+O(H^3),\\
&\varPhi(\bm{R},Z)=\bar{V} Z+\frac{1}{6}\varPhi_3(\bm{R})\Big(Z^3-\frac{H^2}{4}Z\Big)+O(H^4),
\end{split}
\end{align}
where \(\bm r\) and \(\bm b\) are in-plane vector fields,  \(a\) and \(p_0\) are determined from the membrane solution through \eqref{eq:asol} and \eqref{eq:p0}, respectively, and \(c\), \(p_2\), and \(\varPhi_3\) represent higher-order correction fields. The form \(Z^2-H^2/12\) follows from the definition \eqref{eq:rR} of the actual middle surface, and the term \(Z^3-H^2Z/4\) is introduced to satisfy the prescribed voltage condition \eqref{eq:bc} on the upper and lower surfaces.

Using the higher-order expansion, the deformation gradient and electric field are obtained to second order in \(H\) as
\begin{align}\label{eq:FE}
\begin{split}
&\bm{F}=\bm{F}_0+Z\bm{F}_1+\frac{Z^2}{2}\bm{F}_2-\frac{H^2}{24}\nabla\bm{b},\\
&\bm{E}=\bm{E}_0+\frac{1}{2}\varPhi_3\Big(\frac{H^2}{12}-Z^2\Big)\bm{k},
\end{split}
\end{align}
where  $\bm{F}_i$ ($i=0,1,2$) are given by
\begin{align}\label{eq:F}
&\bm{F}_0=\nabla\bm{r}+a\bm{k}\otimes\bm{k},\qquad \bm{F}_1=\bm{k}\otimes\nabla a+\bm{b}\otimes\bm{k},\qquad \bm{F}_2=\nabla\bm{b}+c\bm{k}\otimes\bm{k},
\end{align}
and $\bm{E}_0=-\bar{V}\boldsymbol{k}$ represents the average electric field across the thickness, while the second term describes the higher-order correction associated with the through-thickness variation of the electric field.

Substituting \eqref{eq:FE} into the total potential energy \eqref{eq:TE}, using the expressions for \(a\) and \(p_0\) obtained from the membrane model, and retaining terms up to order \(H^3\), we obtain
\begin{align}\label{eq:ae}
\mathcal{E}[\bm{r},\bm{b}]= H \iint_{\mathcal{R}} \Big(\Omega(\bm{F}_0,\bm{E}_0)+\frac{H^2}{24} \bm{F}_1:\mathcal{A}:\bm{F}_1-
P\bm e_1\cdot\frac{\partial\bm r}{\partial X_1}\Big)\,dA+O(L_1L_2H^4),
\end{align}
where $\mathcal{A}$ is the constrained incremental modulus tensor given by
\begin{align}
\mathcal A=\frac{\partial^2\Omega}{\partial\bm{F}^2}{(\bm{F}_0,\bm{E}_0)}-p_0 \frac{\partial^2 J}{\partial\bm{F}^2}(\bm{F}_0)
\end{align}
with $J=\det(\bm{F})$.

Substituting the expression for \(\bm F_1\) in \eqref{eq:F}, the next-order energy can be written as a quadratic form in \(\bm b\):
\begin{align}\label{eq:qf}
\frac{1}{24}\bm{F}_1:\mathcal{A}:\bm{F}_1=\frac{1}{24}[\bm{A}\bm{b}\cdot\bm{b}+2(\mathcal{A}:\bm{k}\otimes\nabla a)\bm{k}\cdot \bm{b}+\bm{k}\otimes \nabla a:\mathcal{A}:\bm{k}\otimes\nabla a],
\end{align}
where $\bm{A}$ is the acoustic tensor associated with $\mathcal{A}$, defined by
\begin{align}
&\bm{A}\bm{u}:=(\mathcal{A}:\bm{u}\otimes \bm{k})\bm{k},\qquad \forall\, \bm{u}\in\mathbb{R}^3.
\end{align}
Its components with respect to the ordered basis \((\bm e_1,\bm e_2,\bm k)\) are \(A_{ij}=\mathcal A_{i3j3}\). Minimizing \eqref{eq:qf} with respect to $\bm b$ gives \citep{steigmann2007thin,steigmann2007asymptotic,steigmann2008two}
\begin{align}
\bm b
=-\bm A^{-1}
(\mathcal A:\bm k\otimes\nabla a)\bm k.
\end{align}

Substituting this expression for \(\bm b\) into \eqref{eq:ae},  we obtain the plate energy
\begin{align}\label{eq:plate}
\mathcal{E}_\text{pl}[\bm{r}]=\iint_{\mathcal{R}}\Big(G(\nabla\bm{r})+\frac{H^2}{24} \bm{B}\nabla a\cdot\nabla a-P\bm{e}_1\cdot\frac{\partial\bm{r}}{\partial X_1}\Big)\,dA,
\end{align}
where we have used \(G(\nabla\bm r)=\Omega(\bm F_0,\bm E_0)\) and recall that \(a=\det(\nabla\bm r)^{-1}\). The second-order tensor \(\bm B=\bm B(\nabla\bm r)\) is given by
\begin{align}
\begin{split}
&\bm{B}\bm{u}=(\mathcal{A}:\bm{k}\otimes \bm{u})^T\bm{k}-[\bm{A}^{-1}(\mathcal{A}:\bm{k}\otimes\bm{u})\bm{k}\otimes\bm{k}:\mathcal{A}]^T\bm{k},\qquad \forall\, \bm{u}\in\mathbb{R}^3,\\
&B_{ij}=\mathcal{A}_{3i3j}-A^{-1}_{kl}\mathcal{A}_{k33i}\mathcal{A}_{l33j}.
\end{split}
\end{align}
The plate model extends the membrane model by incorporating a strain-gradient term that captures bending effects, and reduces to the membrane model as \(H\to0\).

The equilibrium equations and boundary conditions associated with the plate model follow directly from the variational principle. Their explicit forms are considerably more involved than those of the membrane model and are therefore omitted for brevity.

\section{Homogeneous deformation and the Hessian stability criterion}\label{sec:hom}

We first consider homogeneous deformations, which admit the affine representation
\begin{align}
x_1=\lambda_1 X_1,\qquad x_2=\lambda_2 X_2,\qquad     z=\lambda_1^{-1}\lambda_2^{-1}Z,\qquad E_3=-\frac{V}{H},
\end{align}
where $\lambda_1$ and $\lambda_2$ are the constant stretches in the $X_1$ and $X_2$-directions and the incompressibility constraint has been used. The expansion coefficients in the reduced models are given by
\begin{align}\label{eq:homexp}
\bm{r}=(\lambda_1X_1,\lambda_2X_2,0),\qquad a=\lambda_1^{-1}\lambda_2^{-1},\qquad \bm{b}=\bm{0},\qquad \bm{E}_0=E_3\bm{k}.
\end{align}
Consequently, the membrane approximation \eqref{eq:kinematics} is exact for homogeneous deformations.

Throughout this work,  we adopt the Gent ideal dielectric as in  \cite{huang2012electromechanical}, whose free energy density  is given by
\begin{align}\label{eq:TGent}
\Omega(\bm{F},\bm{E})=-\frac{\mu J_m}{2}\ln\Big(1-\frac{I_1-3}{J_m}\Big)-\frac{1}{2}\epsilon\bm{E}\cdot\bm{C}^{-1}\bm{E},
\end{align}
where $\mu$ denotes the shear modulus, $J_m$ is a dimensionless material parameter that limits the maximum stretch, \(\epsilon\) is the dielectric permittivity of the material, and $I_1=\tr(\bm{C})$ is the first invariant of the right Cauchy-Green tensor $\bm{C}=\bm{F}^T\bm{F}$. In all numerical calculations, the material parameters are taken as
\begin{align}  
\mu=1,\qquad J_m=69.
\end{align}

For homogeneous deformations, the free energy density of the Gent ideal dielectric can be expressed as a function of $\lambda_1$, $\lambda_2$ and $E_3$,
\begin{align}
\Omega(\bm{F},\bm{E})=\bar{\Omega}(\lambda_1,\lambda_2,E_3):=w(\lambda_1,\lambda_2)-\frac{1}{2}\epsilon \lambda_1^2\lambda_2^2 E_3^2,
\end{align}
where $w(\lambda_1,\lambda_2)$ is a reduced strain energy
\begin{align}\label{eq:wre}
w(\lambda_1,\lambda_2)=-\frac{\mu J_m}{2}\ln\Big(1-\frac{\lambda_1^2+\lambda_2^2+\lambda_1^{-2}\lambda_2^{-2}-3}{J_m}\Big).
\end{align}
The  membrane energy \eqref{eq:memcomp} then reduces to
\begin{align}
\mathcal{E}_\text{hom}=\Big(w(\lambda_1,\lambda_2)-\frac{1}{2} \lambda_1^2\lambda_2^2 U^2-P\lambda_1\Big)L_1L_2,
\end{align}
where $U=\sqrt{\epsilon}V/H$ is the scaled voltage.  Requiring \(\mathcal{E}_\text{hom}\) to be stationary with respect to \(\lambda_1\) and \(\lambda_2\) yields
\begin{align}
&w_1(\lambda_1,\lambda_2)-\lambda_1 \lambda_2^2 U^2-P=0, \label{eq:homsol1}\\
&w_2(\lambda_1,\lambda_2)-\lambda_1^2 \lambda_2	U^2=0, \label{eq:homsol2}
\end{align}
where $w_1=\partial w/\partial\lambda_1 $ and $w_2=\partial w/\partial\lambda_2$. Solving for \(P\) and \(U\) gives
\begin{align}\label{eq:hom}
P=\frac{\lambda_1w_1(\lambda_1,\lambda_2)-\lambda_2w_2(\lambda_1,\lambda_2)}{\lambda_1},\qquad U=\sqrt{\frac{w_2(\lambda_1,\lambda_2)}{\lambda_1^2\lambda_2}}.
\end{align}

\begin{figure}[h!]
	\centering
	\subfloat[]{\includegraphics[width=0.44\textwidth]{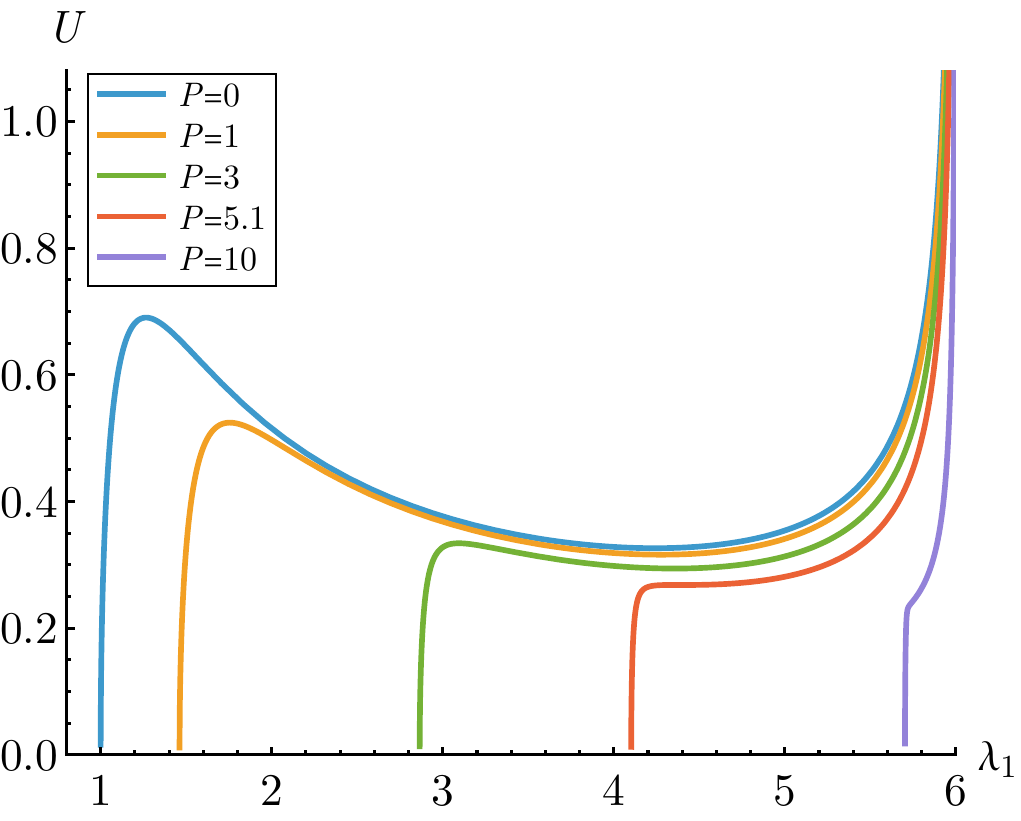}
	}\qquad\quad
	\subfloat[]{\includegraphics[width=0.44\textwidth]{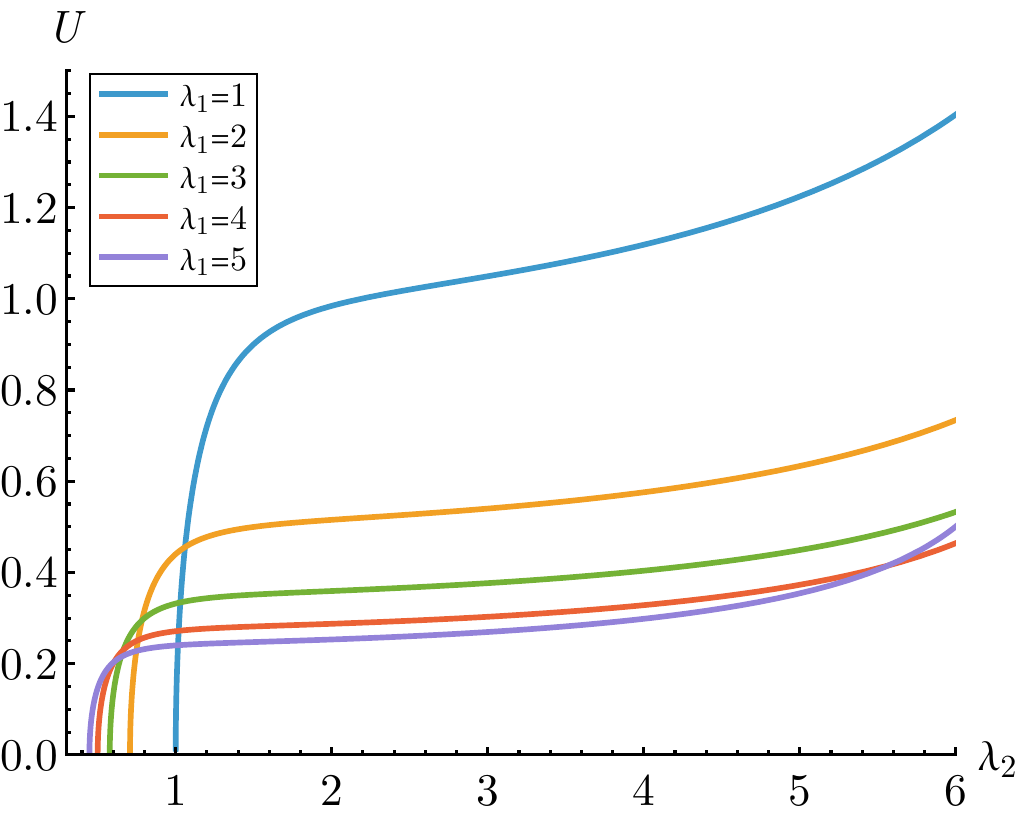}}\\
\hspace{-1.4em}	\subfloat[]{\includegraphics[width=0.43\textwidth]{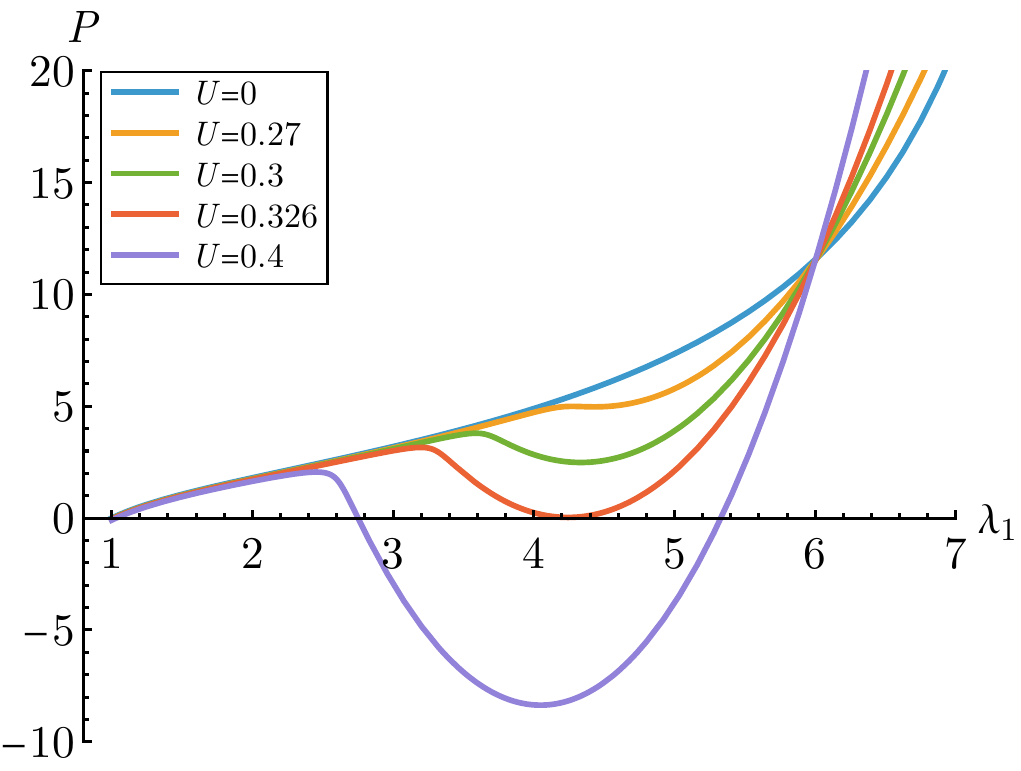}}\qquad\qquad
	 \subfloat[]{\includegraphics[width=0.43\textwidth]{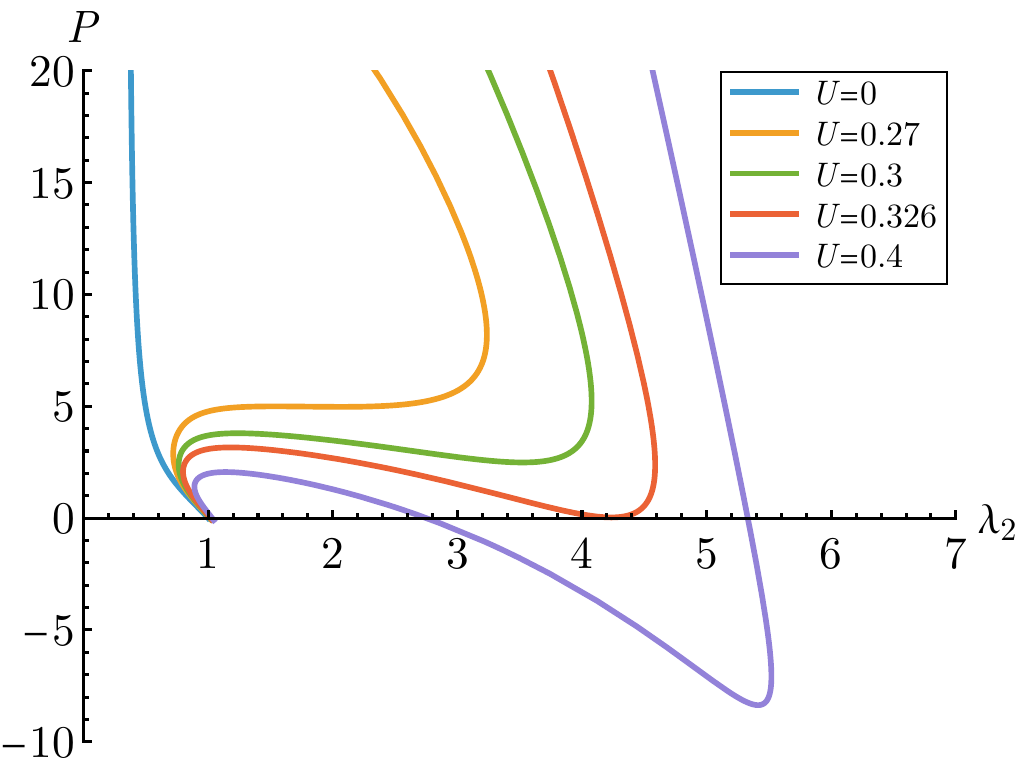}}
	\caption{(a)  \(U\)--\(\lambda_1\) curves for prescribed values of tension; (b) $U$--$\lambda_2$ curves for prescribed values of $\lambda_1$;  (c) \(P\)--\(\lambda_1\) curves for prescribed values of voltage; (d) \(P\)--\(\lambda_2\) curves for prescribed values of voltage.}
	\label{fig:homsol}
\end{figure}

Fig.~\ref{fig:homsol} illustrates the dependence of the loading parameters \(P\) and \(U\) on the principal stretches \(\lambda_1\) and \(\lambda_2\), where \(\lambda_1\) denotes the stretch along the longitudinal direction and \(\lambda_2\) represents the transverse stretch. For a prescribed tension \(P\), the voltage--longitudinal stretch relation shown in Fig.~\ref{fig:homsol}(a) undergoes a transition from non-monotonic to monotonic behavior as \(P\) increases. When \(P\) is small, multiple longitudinal stretch states coexist for a given voltage. This multiplicity disappears with increasing \(P\), and the response becomes monotonic for \(P\geq 5.1\). The voltage--transverse stretch relation is further shown in Fig.~\ref{fig:homsol}(b) for prescribed values of \(\lambda_1\). Unlike the voltage--longitudinal stretch relations , these curves remain monotonic over the range considered, indicating that no limit-point behavior appears along this loading path.

Under prescribed voltage, the tension--stretch responses along the longitudinal and transverse directions are shown in Fig.~\ref{fig:homsol}(c) and (d). For small values of \(U\), both the longitudinal and transverse responses remain monotonic. As \(U\) increases, non-monotonic regions develop in the tension--longitudinal stretch relation, and a turning point appears (e.g., at \(U=0.27\)), indicating the loss of monotonicity of the homogeneous response. For larger values of \(U\) (e.g., \(U=0.326\)), the required tension \(P\) becomes negative over a range of stretches. In this regime, compressive stresses are required to maintain the homogeneous deformation, suggesting the possible emergence of alternative deformation modes, such as wrinkling.

The stability of homogeneous deformations can be assessed using the Hessian criterion \citep{zhao2007method}. To state this criterion, we introduce the electric displacement
\begin{align}
D_3= -\frac{\partial \bar{\Omega}}{\partial E_3}=\epsilon\lambda_1^2\lambda_2^2 E_3,
\end{align}
and define the Legendre-transformed free energy density
\begin{align}\label{eq:Lg}
\Omega^*(\lambda_1,\lambda_2,D_3)
= \bar{\Omega}(\lambda_1,\lambda_2,E_3) + E_3 D_3=w(\lambda_1,\lambda_2)+\frac{1}{2} \frac{D_3^2}{\epsilon\lambda_1^2\lambda_2^2}.
\end{align}
The Hessian criterion requires  the Hessian matrix of $\Omega^*$
\begin{align}\label{eq:Hst}
H_{\Omega^*}=\begin{pmatrix}
\frac{\partial^2\Omega^*}{\partial\lambda_1^2} & \frac{\partial^2\Omega^*}{\partial\lambda_1\partial\lambda_2} & \frac{\partial^2\Omega^*}{\partial\lambda_1\partial D_3}\\
\frac{\partial^2\Omega^*}{\partial\lambda_2\partial\lambda_1} & \frac{\partial^2\Omega^*}{\partial\lambda_2^2} & \frac{\partial^2\Omega^*}{\partial\lambda_2\partial D_3}\\
\frac{\partial^2\Omega^*}{\partial D_3\partial\lambda_1} & \frac{\partial^2\Omega^*}{\partial D_3\partial\lambda_2} & \frac{\partial^2\Omega^*}{\partial D_3^2}\\
\end{pmatrix},
\end{align}
to be positive definite for stability. A necessary condition for instability is therefore
\begin{align}\label{eq:H0}
\det(H_{\Omega^*})=0.
\end{align}
Using \eqref{eq:Lg} and \eqref{eq:Hst}, this condition becomes
\begin{align}\label{eq:Hess}
\begin{split}
&3 \lambda_2^2+\lambda_1^2\big(1-6(3+J_m)\lambda_2^4+2\lambda_2^6\big)-\lambda_1^6\lambda_2^2(2+7\lambda_2^6)\\
&-\lambda_1^4\lambda_2^2\big(2(3+J_m)-22\lambda_2^2-2(3+J_m)\lambda_2^6+\lambda_2^8\big)=0.
\end{split}
\end{align}
For the loading paths considered here, the homogeneous deformation evolves continuously from the undeformed stable state, where the Hessian matrix is positive definite. Therefore, the first loss of positive definiteness occurs when the smallest eigenvalue of the Hessian reaches zero, which is detected by the condition \eqref{eq:H0}.

The Hessian criterion only determines the onset of instability and does not specify the deformation that emerges beyond this point. Determining the post-instability deformation requires a separate bifurcation analysis.

\section{Linear bifurcation analysis}\label{sec:lin}

We now investigate the loss of stability of the homogeneous deformation through a linear bifurcation analysis based on the membrane model. We first derive the dispersion relation governing infinitesimal perturbations of the homogeneous state and then establish its connection with the Hessian stability criterion introduced in the previous section.

\subsection{Dispersion relation}

To analyze the bifurcation of the homogeneous deformation, we introduce a perturbed solution of the form
\begin{align}
r_1(X_1,X_2)=\lambda_1 X_1+u(X_1,X_2),\quad r_2(X_1,X_2)=\lambda_2 X_2+v(X_1,X_2),
\end{align}
where $u$ and $v$ are small perturbations in the $X_1$- and $X_2$-directions. We seek normal-mode solutions  of the form
\begin{align}\label{eq:uvlinear}
u(X_1,X_2)=f(X_2) e^{\ri kX_1},\quad v(X_1,X_2)=g(X_2)\ri ke^{\ri kX_1},
\end{align}
where $k$ is the longitudinal wavenumber and $f$ and $g$ are functions to be determined. This ansatz describes perturbations  periodic in the loading direction \(X_1\). The factor \(\ri k\) in the expression for \(v\) is introduced for convenience so that the resulting equations remain real.

Substituting \eqref{eq:uvlinear} into the equilibrium equations \eqref{eq:equi} and the boundary conditions \eqref{eq:bcc1}--\eqref{eq:bcc2}, and then linearizing about the homogeneous state, we obtain a linear boundary value problem for \(f\) and \(g\). Introducing the state vector \(\bm y=(f,f',g,g')^T\), this problem can be written in the first-order form
\begin{align}
&\frac{d\bm{y}}{dX_2}=M(k)\bm{y},\label{eq:y}\\
&N(k)\bm{y}=0,\quad \text{on}\ X_2=\pm\frac{L_2}{2}. \label{eq:ybc}
\end{align}
 The coefficient matrices \(M\) and \(N\) are given by
\begin{align}\label{eq:coeM}
M=\begin{pmatrix}
0 & 1 & 0 & 0\\
m_{21} & 0 & 0 & m_{24}\\
0  & 0 & 0 & 1\\
0  & m_{42} & m_{43} & 0
\end{pmatrix},\quad N=\begin{pmatrix}
0 & n_{12} & n_{13} & 0\\
n_{21} & 0 & 0 & n_{24}
\end{pmatrix},
\end{align}
where the coefficients depend on the homogeneous state and the wavenumber \(k\), but are constant for fixed values of these quantities. Their explicit expressions are listed in \ref{app:coe}. The above formulation is reminiscent of the Stroh formalism widely used in incremental elasticity \citep{stroh1962steady,ting1996anisotropic}, although the matrix \(M\) does not possess the canonical Stroh structure in the present problem.

The  boundary value problem \eqref{eq:y}--\eqref{eq:ybc} can be solved as follows. Let \(Q\) be a \(4\times2\) matrix whose columns span the null space of \(N\), so that 
\begin{align}
NQ=0.
\end{align}
The columns of \(Q\) represent two linearly independent solutions satisfying the boundary conditions at \(X_2=-L_2/2\). Since the matrix \(M\) is independent of \(X_2\), the solution of the first-order linear system \eqref{eq:y} can be expressed using the matrix exponential as
\begin{align}
\bm y(X_2)=e^{(X_2+L_2/2)M}\bm y(-L_2/2).
\end{align}
Therefore, evaluating the two independent solutions satisfying the boundary conditions at \(X_2=-L_2/2\) at \(X_2=L_2/2\) gives the solution matrix \(e^{L_2M}Q\). Imposing the boundary conditions at \(X_2=L_2/2\), the existence of a non-trivial solution requires
\begin{align}\label{eq:de}
E(k,\lambda_1,\lambda_2):=\det(Ne^{L_2 M}Q)=0,
\end{align}
which defines the dispersion relation of the bifurcation. The explicit expression for $E$ is available but omitted for brevity.

Owing to the symmetry of the domain and boundary conditions about \(X_2=0\), the dispersion relation decomposes into symmetric and antisymmetric branches. In the symmetric branch, \(u\) is an even function of \(X_2\), whereas \(v\) is an odd function of \(X_2\). The resulting deformation is therefore symmetric with respect to the midline \(X_2=0\). In the antisymmetric branch, the parity is reversed: \(u\) is odd and \(v\) is even in \(X_2\), giving rise to deformations  that are antisymmetric about \(X_2=0\).

We consider two typical loading scenarios. In the first, the applied tension $P$ is fixed while the voltage is increased from zero \citep{zhao2007method}. In the second, the membrane is first stretched to a prescribed stretch $\lambda_1$, which is subsequently maintained while the voltage is increased from zero \citep{huang2012giant}. In the numerical calculations,  the width is normalized such that $L_2=1$.

\begin{figure}[h!]
	\centering
	\subfloat[]{\includegraphics[width=0.435\textwidth]{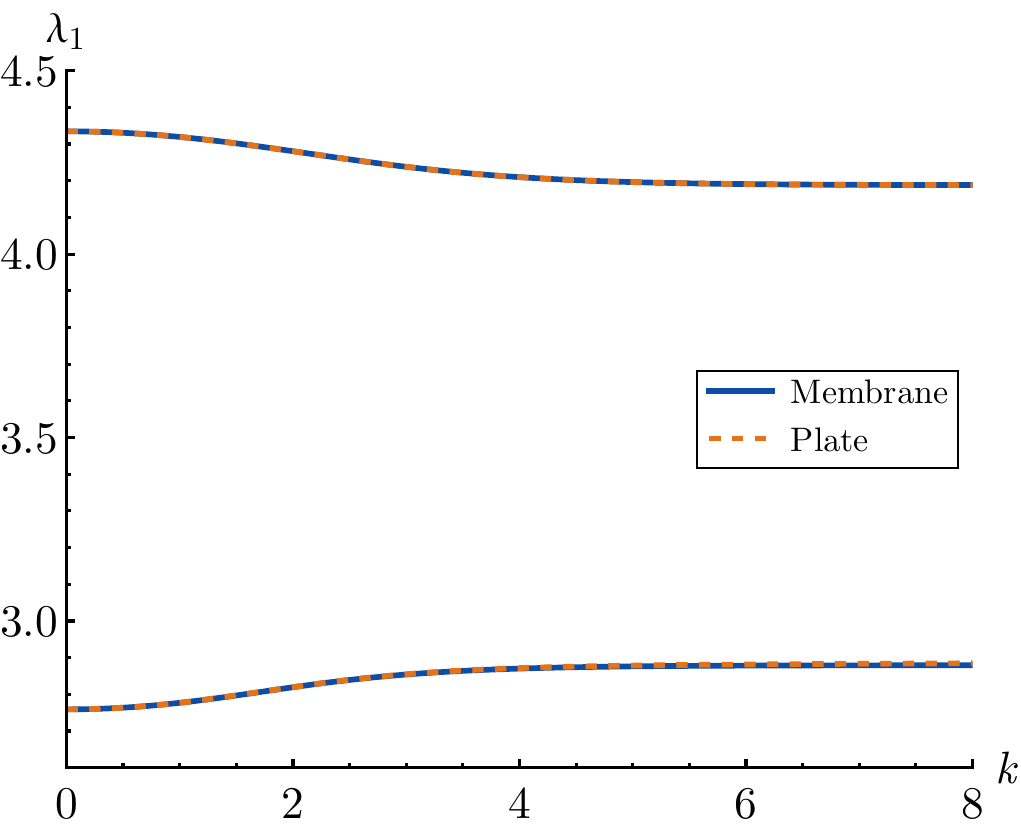}
	}\qquad\qquad
	\subfloat[]{\includegraphics[width=0.44\textwidth]{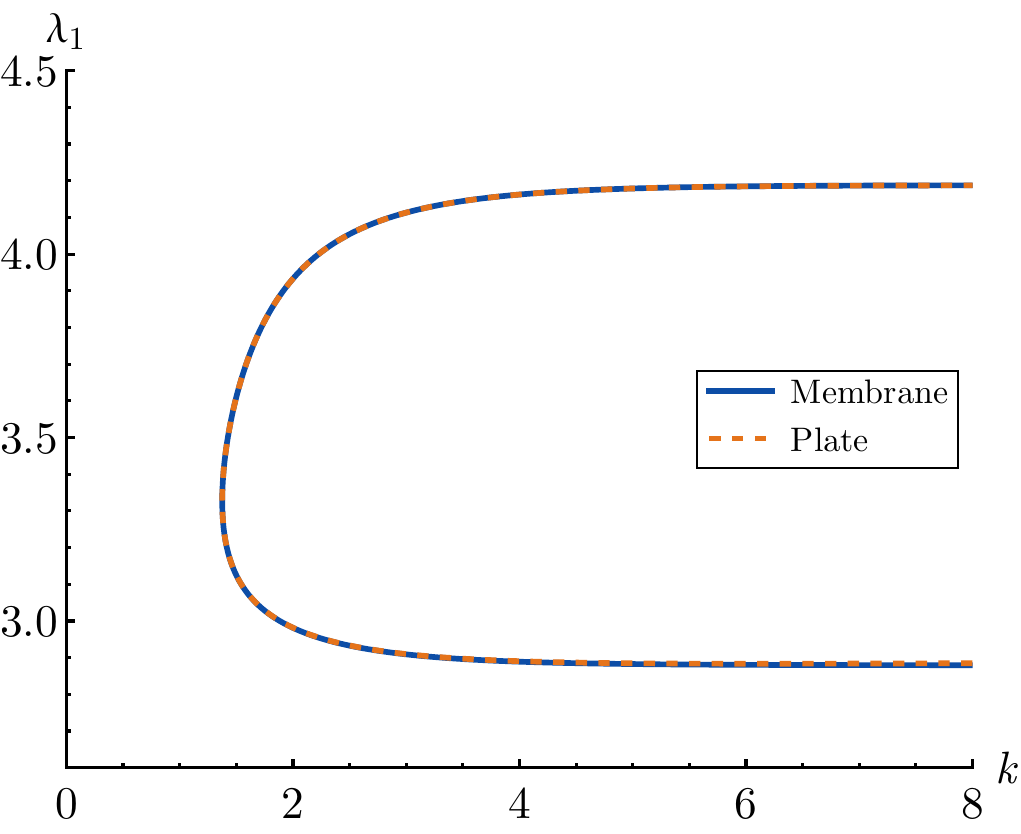}
	}
	\caption{Dispersion curves for a fixed tension \(P=2.5\) predicted by the membrane and plate models: (a) symmetric branch; (b) antisymmetric branch.}
	\label{fig:dp1}
\end{figure}

\begin{figure}[h!]
	\centering
	\subfloat[]{\includegraphics[width=0.435\textwidth]{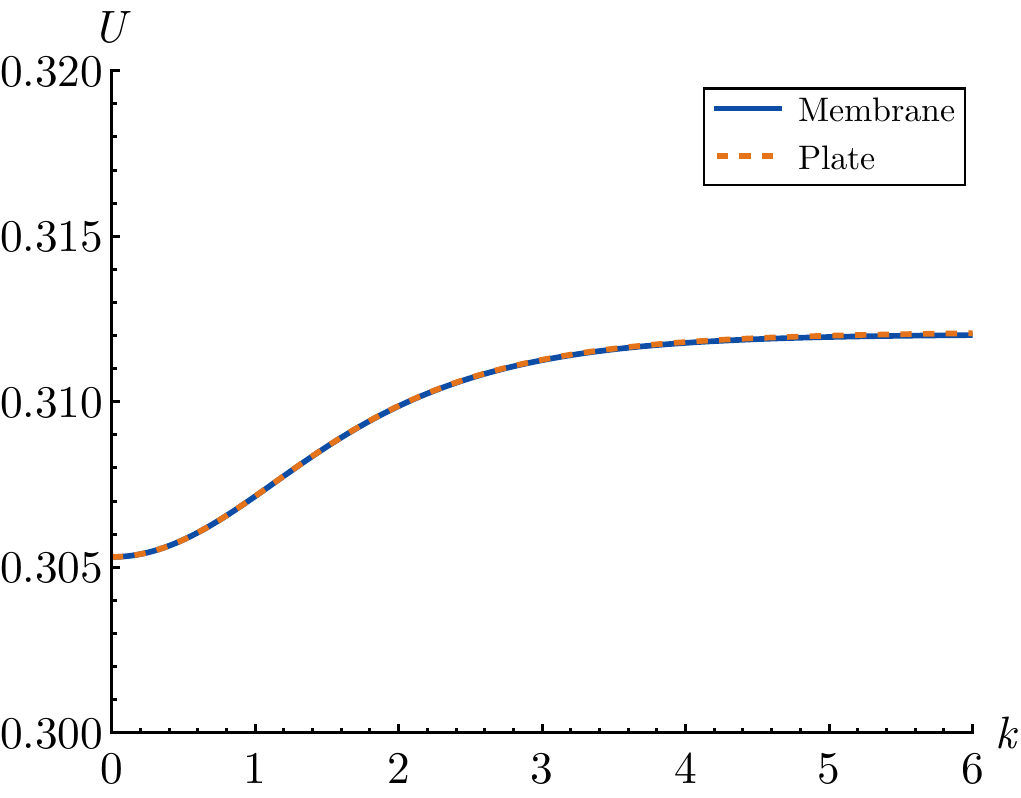}
	}\qquad\qquad
	\subfloat[]{\includegraphics[width=0.44\textwidth]{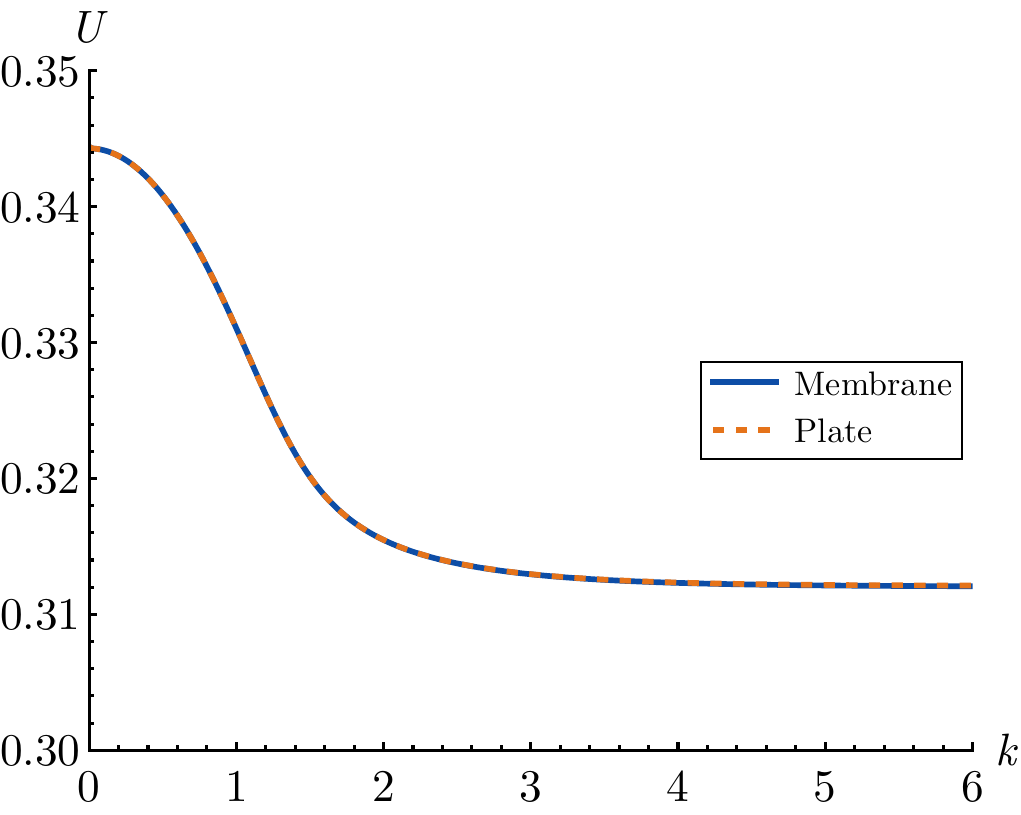}
	}
	\caption{Dispersion curves  for a fixed  stretch \(\lambda_1=3.5\) predicted by the membrane and plate models: (a) symmetric branch; (b) antisymmetric branch.}
	\label{fig:dp2}
\end{figure}

Figures~\ref{fig:dp1} and \ref{fig:dp2} show the dispersion curves predicted by the membrane and plate models for a fixed tension \(P=2.5\) and a fixed stretch \(\lambda_1=3.5\), respectively. The membrane results are obtained from \eqref{eq:de}, while  the plate results are obtained from the corresponding bifurcation condition derived from the higher-order plate model. For the plate model, a thickness ratio \(H/L_2=0.2\) is used. In both loading scenarios, the minimum load occurs at the critical wavenumber \(k=0\), and the dispersion curves predicted by the membrane and plate models are nearly indistinguishable. These observations indicate that bending stiffness has little influence on the linear bifurcation behavior and that the membrane model captures the essential features of the instability. The vanishing critical wavenumber further implies that linear analysis alone cannot determine the resulting deformation mode, and that nonlinear effects are required to resolve it \citep{fu2001nonlinear}.

\subsection{Connection with the Hessian stability criterion}\label{sec:exp}

Having confirmed that  the bifurcation occurs at a zero wavenumber, we now employ a perturbation approach to derive an explicit analytical condition characterizing its onset.

When $k$ is small, we expand the coefficient matrices $M(k)$ and $N(k)$ in the form
\begin{align}
M(k)=M_0+k^2 M_1+O(k^4),\quad N(k)=N_0+k^2 N_1+O(k^4).
\end{align}
This expansion is readily obtainable  from \eqref{eq:coeM} and reflects the fact that $k$ enters the linearized equations only through $k^2$. Accordingly,  the matrix \(Q(k)\) spanning the null space of \(N(k)\) admits the expansion
\begin{align}
Q(k)=Q_0+k^2 Q_1+O(k^4).
\end{align}
Using Duhamel's formula, the matrix exponential admits the expansion
\begin{align}
e^{L_2 M(k)}=e^{ L_2 M_0}+k^2 L_2\int_0^1 e^{(1-s)L_2 M_0} M_1 e^{s L_2 M_0}\,ds+O(k^4).
\end{align}
The leading-order term $e^{L_2 M_0}$ can be evaluated explicitly, since $M_0$ is nilpotent with $M_0^3=0$. Hence
\begin{align}
e^{L_2 M_0}=I+L_2 M_0+\frac{L_2^2}{2}  M_0^2.
\end{align}

Substituting the above asymptotic expansions into \eqref{eq:de}, we obtain  an asymptotic expansion of $E(k,\lambda_1,\lambda_2)$ of the form
\begin{align}\label{eq:Ek}
\begin{split}
E(k,\lambda_1,\lambda_2)=&k^4 \Phi(\lambda_1,\lambda_2)\big[3 \lambda_2^2+\lambda_1^2\big(1-6(3+J_m)\lambda_2^4+2\lambda_2^6\big)-\lambda_1^6\lambda_2^2(2+7\lambda_2^6)\\
&-\lambda_1^4\lambda_2^2\big(2(3+J_m)-22\lambda_2^2-2(3+J_m)\lambda_2^6+\lambda_2^8\big)\big]+O(k^6),
\end{split}
\end{align}
where $\Phi(\lambda_1,\lambda_2)$ is the rational function given by
\begin{align}
\begin{split}
& \Phi(\lambda_1,\lambda_2)=\frac{
\mu^2	J_m^2 L_2^2  
	\lambda_1 \lambda_2^3 (\lambda_1^2-\lambda_2^2)
}{
	\big(
	1+\lambda_1^4\lambda_2^2
	+\lambda_1^2\lambda_2^2(\lambda_2^2-J_m-3)
	\big)^2\Psi(\lambda_1,\lambda_2)
},
\end{split}
\end{align}
in which
\begin{align}
& \Psi(\lambda_1,\lambda_2)=2\lambda_1^6\lambda_2^6
+2\lambda_1^2\lambda_2^2(J_m+3-\lambda_2^2)
+\lambda_1^4\lambda_2^2\big(\lambda_2^6-(J_m+3)\lambda_2^4-3\big)-1
.
\end{align}
The absence of lower-order terms (i.e., $k^0$ and $k^2$) in \eqref{eq:Ek} follows from the identity 
\begin{align}
N_0e^{L_2 M_0}Q_0=0,
\end{align}
implying that \(k=0\) is always a root of multiplicity at least four.

The onset of bifurcation is therefore associated not with the existence of the trivial root \(k=0\), but with a change in its multiplicity. A necessary condition for the existence of nontrivial solutions with arbitrarily small wavenumbers is that the leading-order coefficient vanishes. Indeed, if it is nonzero, then $E(k,\lambda_1,\lambda_2)\neq 0$ for sufficiently small $k$, and no nearby nontrivial root exists. From \eqref{eq:Ek} and noting that 
$\Phi(\lambda_1,\lambda_2)$ is generically nonzero, the bifurcation condition reduces to
\begin{align}\label{eq:bif}
\begin{split}
&3 \lambda_2^2+\lambda_1^2\big(1-6(3+J_m)\lambda_2^4+2\lambda_2^6\big)-\lambda_1^6\lambda_2^2(2+7\lambda_2^6)\\
&-\lambda_1^4\lambda_2^2\big(2(3+J_m)-22\lambda_2-2(3+J_m)\lambda_2^6+\lambda_2^8\big)=0.
\end{split}
\end{align}
This condition coincides exactly with the Hessian  stability criterion \eqref{eq:Hess}, establishing that Hessian instability corresponds to a zero-wavenumber bifurcation.

\section{Weakly nonlinear analysis}\label{sec:weak}

In this section, we carry out a weakly nonlinear analysis to determine the deformation mode associated with the zero-wavenumber bifurcation. 

For definiteness, we focus on the case of fixed applied tension $P$; the alternative loading case of fixed $\lambda_1$ can be addressed  by a similar procedure. We take $\lambda_1$ as the control parameter and denote its critical value at bifurcation by $\lambda_{1\text{cr}}$. In the near-critical region, we write
\begin{align}
\lambda_1=\lambda_{1\text{cr}}+\varepsilon \lambda_0,
\end{align}
where $\lambda_0$ is an $O(1)$ constant and $\varepsilon$ is a positive parameter measuring the deviation from the critical state. The associated transverse stretch $\lambda_2$ is expanded as
\begin{align}
\lambda_2=\lambda_{2\text{cr}}+\varepsilon d_1+\varepsilon^2 d_2+\varepsilon^3 d_3+\cdots,
\end{align}
where the coefficients \(d_i\) are determined by enforcing the condition that \(P\) remains fixed.

Since the dispersion equation \eqref{eq:de} depends on $k^2$, it follows that the bifurcation mode satisfies $k=O(\varepsilon^{1/2})$. Consequently, the spatial dependence of the near-critical solution on the longitudinal coordinate $  X_1 $ is captured by the stretched variable
\begin{align}
S=\varepsilon^{1/2}X_1.
\end{align}
The relative orders of $u$ and $v$ are deduced by expanding the linear solution \eqref{eq:uvlinear} for small $k$, while their absolute sizes are determined by the fact that the amplitude is expected to vary linearly with $\lambda_1-\lambda_{1\text{cr}}$ for the bifurcations under consideration. This scaling analysis leads to $u=O(\varepsilon^{1/2})$ and $v=O(\varepsilon)$. Accordingly, we look for a near-critical solution of the form
\begin{align}
\begin{split}\label{eq:asy}
&r_1(X_1,X_2)=\lambda_1 X_1+\varepsilon^{1/2}u_1(S,X_2)+\varepsilon^{3/2}u_2(S,X_2)+\varepsilon^{5/2}u_3(S,X_2)+\cdots,\\
&r_2(X_1,X_2)=\lambda_2 X_2+\varepsilon v_1(S,X_2)+\varepsilon^{2}v_2(S,X_2)+\varepsilon^{3}v_3(S,X_2)+\cdots,
\end{split}
\end{align}
where all the functions on the right-hand side are to be determined order by order. Substituting this asymptotic solution into the equilibrium equations \eqref{eq:equi} as well as the boundary conditions \eqref{eq:bcc1}--\eqref{eq:bcc2}, and collecting terms at each order of $\varepsilon$, yields a hierarchy of boundary value problems.

\medskip
\noindent\textit{Leading-order problem.}
At leading order, we obtain a homogeneous linear system for $u_1$ and $v_1$:
\begin{align}
&\frac{\partial^2 u_1}{\partial X_2^2}=0,\quad \frac{\partial^2 v_1}{\partial X_2^2}+\alpha_1 \frac{\partial^2 u_1}{\partial S\partial X_2}=0,\label{eq:ll}\\
&\frac{\partial u_1}{\partial X_2}=0,\quad  \frac{\partial v_1}{\partial X_2}+\beta_1\frac{\partial u_1}{\partial S}=0,\qquad  \ X_2=\pm\frac{L_2}{2},\label{eq:bcll}
\end{align}
where $\alpha_1$ and $\beta_1$ are constants  available but omitted for brevity. The  solution to  \eqref{eq:ll} and \eqref{eq:bcll} is readily obtained as
\begin{align}\label{eq:lead}
u_1=A(S),\quad v_1=c_1(S)X_2+c_2(S),
\end{align}
where $c_1(S)=-\beta_1 A'(S)$, and $A(S)$ and $c_2(S)$ are arbitrary functions. In particular, the leading-order solution is fully determined by the amplitude function $A(S)$ and an undetermined function $c_2(S)$.

\medskip
\noindent\textit{Second-order problem.} At the next order, the governing equations become inhomogeneous and take the form
\begin{align}
&\frac{\partial^2 u_2}{\partial X_2^2}=\alpha_2 A''(S), \label{eq:s1}\\
& \frac{\partial^2 v_2}{\partial X_2^2}+\alpha_3 \frac{\partial^2 u_2}{\partial S\partial X_2}=\alpha_4 X_2 A^{(3)}(S)+\alpha_5 c_2''(S),\label{eq:s2}\\
&\frac{\partial u_2}{\partial X_2}=\beta_2 X_2 A''(S)+\beta_3 c_2'(S),\qquad  \ X_2=\pm\frac{L_2}{2},\label{eq:bcs1}\\
&\frac{\partial v_2}{\partial X_2}+\beta_4\frac{\partial u_2}{\partial S\partial X_2}=\beta_5 A'(S)+\beta_6 A'(S)^2,\qquad \ X_2=\pm\frac{L_2}{2},\label{eq:bcs2}
\end{align}
where $\alpha_i$ and $\beta_i$ are constants. By direct integration, the general solution to \eqref{eq:s1} and \eqref{eq:s2} can be written as
\begin{align}\label{eq:second}
\begin{split}
&u_2=\frac{1}{2}\alpha_2 A''(S) X_2^2 +c_3(S)X_2,\\ 
&v_2=\frac{1}{6}(\alpha_4-\alpha_2\alpha_3)A^{(3)}(S)X_2^3+\frac{1}{2}(\alpha_5 c_2''(S)-\alpha_3 c_3'(S))X_2^2+c_4(S)X_2+c_5(S),
\end{split}
\end{align}
where $c_3(S)$, $c_4(S)$ and $c_5(S)$ are arbitrary functions. The integration function depending only on $S$ in $u_2$ has been omitted, as it can be absorbed into the leading-order solution.

 Imposing the boundary conditions \eqref{eq:bcs1} and \eqref{eq:bcs2} at $X_2=-L_2/2$, we obtain
\begin{align}
&c_3(S)=\frac{L_2}{2}(\alpha_2-\beta_2)A''(S)+\beta_3 c'_2(S),\\
&c_4(S)=\gamma_1 A^{(3)}(S)+\gamma_2 c''_2(S)+\beta_5 A'(S)+\beta_6 A'(S)^2,
\end{align}
where $\gamma_1$ and $\gamma_2$ are constants given by
\begin{align}
&\gamma_1=-\frac{L_2^2}{8}\big(\alpha_4+\alpha_2(\alpha_3-\beta_4)+2\beta_2(-\alpha_3+\beta_4)\big),\\
& \gamma_2=\frac{L_2}{2}\big(\alpha_5+\beta_3(\beta_4-\alpha_3)\big).
\end{align}
Then enforcing the boundary conditions at $X_2=L_2/2$  leads to two solvability conditions:
\begin{align}
\alpha_2-\beta_2=0,\qquad c_2''(S)=\frac{L_2}{2}\frac{(\alpha_2-\beta_2)(\alpha_3-\beta_4)}{(\alpha_5-\alpha_3\beta_3+\beta_3\beta_4)}A^{(3)}(S).
\end{align}
It is checked that the first condition recovers the linear bifurcation condition \eqref{eq:bif}, while the second implies
\begin{align}
c_2''(S)=0.
\end{align}
The remaining freedom in $ c_2(S)$ reflects the translational invariance of the problem in the $X_2$-direction. Without loss of generality, this invariance may be removed by setting $c_2(S)=0$.

\medskip
\noindent\textit{Third-order problem.} At third order, only the equations for $u_3$
are required to derive the solvability condition. These equations take the form
\begin{align}
&\frac{\partial^2 u_3}{\partial X_2^2}=h_1,\\
& \frac{\partial u_3}{\partial X_2}=h_2,\qquad \ X_2=\pm\frac{L_2}{2},
\end{align}
where the inhomogeneous terms $h_1$ and $h_2$ on the right-hand sides involve only the leading- and second-order solutions. For a solution to exist, the following solvability condition must hold:
\begin{align}\label{eq:sc}
\int_{-L_2/2}^{L_2/2} h_1\,dX_2=h_2|_{X_2=-L_2/2}^{X_2=L_2/2}.
\end{align}
Substituting the lower-order equations \eqref{eq:lead} and \eqref{eq:second} into \eqref{eq:sc}, after simplification, we obtain an amplitude equation that governs the near-critical behavior:
\begin{align}\label{eq:amp0}
A^{(4)}(S)=\lambda_0 k_1 A''(S)+2 k_2 A'(S)A''(S),
\end{align}
where the coefficients $k_1$ and $k_2$ can be expressed in closed analytical form in terms of the material and geometric parameters, but their explicit expressions are omitted here for brevity.

Assuming that the membrane is infinitely long in the \(X_1\)-direction, we seek solutions that approach the homogeneous state in the far field, i.e., $ \lim_{S\to\pm\infty} A'(S)\to 0$. Integrating \eqref{eq:amp0} once yields 
\begin{align}\label{eq:weak1}
A^{(3)}(S)=\lambda_0 k_1 A'(S)+k_2 A'(S)^2.
\end{align}
It is found that the coefficients satisfy the relation
\begin{align}
k_1=2k_2,
\end{align}
which follows from the symmetry of the two homogeneous solutions in the vicinity of the bifurcation point \citep{ye2020weakly}. Eq. \eqref{eq:weak1} admits a localized solution of the form
\begin{align}
A'(S)=-\frac{3\lambda_0k_1}{2k_2}\sech^2\Big(\frac{\sqrt{\lambda_0k_1}}{2}S\Big),
\end{align}
provided $\lambda_0k_1>0$, with its maximum at \(S=0\).  Integrating once more yields
\begin{align}
A(S)=-\frac{3\sqrt{\lambda_0 k_1}}{k_2}\tanh\Big(\frac{\sqrt{\lambda_0k_1}}{2}S\Big).
\end{align}
 Substituting this result into \eqref{eq:asy}, the leading-order position of the bifurcated solution can be reconstructed as
\begin{align}\label{eq:asy1}
\begin{split}
&x_1=(\lambda_{1\text{cr}}+\varepsilon \lambda_0) X_1+ \varepsilon^{1/2}A(\varepsilon^{1/2} X_1),\\
&x_2=(\lambda_{2\text{cr}}+\varepsilon d_1 -\varepsilon \beta_1 A'(\varepsilon^{1/2}X_1))X_2,\\
&z=
\Big(\frac{1}{\lambda_{1\text{cr}}\lambda_{2\text{cr}}}+\varepsilon\frac{(\beta_1\lambda_{1\text{cr}}-\lambda_{2\text{cr}})A'(\varepsilon^{1/2}X_1)-d_1\lambda_{1\text{cr}}-\lambda_0\lambda_{2\text{cr}}}{\lambda_{1\text{cr}}^2\lambda_{2\text{cr}}^2}\Big)Z,
\end{split}
\end{align}
where we have used $x_1$ and $x_2$ in place of $r_1$  and $r_2$, and the expression for $z$ follows from the incompressibility constraint. In particular, the deformation gradient associated with  \eqref{eq:asy1} remains diagonal  and depends only on $X_1$.

For the alternative loading scenario in which $\lambda_1$ is fixed, we instead use $\lambda_2$ as the control parameter and write
\begin{align}
\lambda_2=\lambda_{2\text{cr}}+\varepsilon \lambda_0,
\end{align}
in the near-critical region, where $\lambda_0$ is an $O(1)$ constant.  Following a similar procedure, we finally arrive at an amplitude equation taking the same form as \eqref{eq:weak1}, but with  different expressions for $k_1$ and $k_2$. This equation describes the near-critical behavior of the bifurcated solution under fixed longitudinal  stretch.

Tables~\ref{tab1} and \ref{tab2} list the linear and nonlinear coefficients in the amplitude equation for the Gent model under the loading conditions \(P=2.5\) and \(\lambda_1=3.5\), respectively. According to  \eqref{eq:bif}, bifurcation occurs at the critical stretches \((\lambda_{\mathrm{1cr}},\lambda_{\mathrm{2cr}})=(2.759,1.148)\) for the former case and \((3.5,1.221)\) for the latter. As shown in  Tables~\ref{tab1} and \ref{tab2}, the coefficients \(k_1\) and \(k_2\) are negative in all cases. It follows that \(\lambda_0<0\), which implies \(A'(S)>0\). The bifurcation is therefore subcritical. The associated deformation is characterized by localized in-plane stretching accompanied by localized thinning in the thickness direction.

Furthermore, the coefficients predicted by the membrane and plate models are in excellent agreement over the range of thickness ratios considered. In particular, for \(H/L_2\leq0.2\), the relative differences in both \(k_1\) and \(k_2\) remain below \(0.5\%\), and are still below \(1\%\) even for \(H/L_2=0.3\). This indicates that bending stiffness has only a negligible effect on the weakly nonlinear coefficients.

Overall, the weakly nonlinear analysis demonstrates that Hessian instability gives rise to a subcritical localized necking mode. The membrane model provides an accurate description not only at the linear bifurcation level but also in the weakly nonlinear regime, while the plate model introduces only small higher-order corrections.

\begin{table}[h!]
	\centering
	\caption{Linear and nonlinear coefficients in \eqref{eq:weak1} given by the membrane model and plate model for fixed $P=2.5$, where $\eta=H/L_2$ denotes the thickness-to-width ratio. The coefficients  $k_1$ and $k_2$ correspond to the membrane model, while $k_{1\text{pl}}$ and  $k_{2\text{pl}}$ are computed from the plate model.}
	\begin{tabular}{lcccccc}
		\hline 
		& $k_1$   & $k_{1\text{pl}}$  &  $|1-k_{1\text{pl}}/k_1|$  & $k_2$ & $k_{2\text{pl}}$  & $|1-k_{2\text{pl}}/k_2|$\\ 
		\hline 
		$\eta=0$	& $-54.502$  & $-54.502$ & \hspace{-1.4em}$0\%$  & $-27.251$ & $-27.251$ & \hspace{-1.4em} $0\%$ \\ 
		\hline 
		$\eta=0.05$	& $-54.502$ & $-54.486$  & $0.03\%$   &   $-27.251$ &  $-27.243$ & $0.03\%$  \\ 
		\hline
		$\eta=0.1$ & $-54.502$ & $-54.440$  & $0.11\%$  &     $-27.251$  & $-27.220$ & $0.11\%$ \\
		\hline 
		$\eta=0.15$ & $-54.502$ & $-54.362$  & $0.26\%$  &     $-27.251$  & $-27.181$ & $0.26\%$ \\
		\hline 
		$\eta=0.2$ & $-54.502$ & $-54.254$  & $0.46\%$  &     $-27.251$  & $-27.127$ & $0.46\%$ \\
		\hline 
		$\eta=0.3$ & $-54.502$ & $-53.947$  & \hspace{-1.4em}$1\%$  &     $-27.251$  & $-26.937$ &  \hspace{-1.4em}$1\%$ \\
		\hline 
	\end{tabular}
	\label{tab1}
\end{table}

\begin{table}[h!]
	\centering
	\caption{Linear and nonlinear coefficients in \eqref{eq:weak1} given by the membrane model and plate model for fixed $\lambda_1=3.5$, where $\eta=H/L_2$ denotes the thickness-to-width ratio. The coefficients  $k_1$ and $k_2$ correspond to the membrane model, while $k_{1\text{pl}}$ and  $k_{2\text{pl}}$ are computed from the plate model.}
	\begin{tabular}{lcccccc}
		\hline 
		& $k_1$   & $k_{1\text{pl}}$  &  $|1-k_{1\text{pl}}/k_1|$  & $k_2$ & $k_{2\text{pl}}$  & $|1-k_{2\text{pl}}/k_2|$\\ 
		\hline 
		$\eta=0$	& $-11.613$  & $-11.613$ & \hspace{-1.4em}$0\%$  & $-14.125$ & $-14.125$ & \hspace{-1.4em} $0\%$ \\ 
		\hline 
		$\eta=0.05$	& $-11.613$ & $-11.611$  & $0.01\%$   &   $-14.125$  &  $-14.123$ & $0.01\%$  \\ 
		\hline
		$\eta=0.1$ &  $-11.613$ & $-11.607$  & $0.04\%$  &     $-14.125$  & $-14.118$ & $0.04\%$ \\
		\hline 
		$\eta=0.15$ &  $-11.613$ & $-11.600$  & $0.11\%$  &      $-14.125$  & $-14.110$ & $0.11\%$ \\
		\hline 
			$\eta=0.2$ &  $-11.613$ & $-11.591$  & $0.19\%$  &      $-14.125$  & $-14.098$ & $0.19\%$ \\
		\hline 
			$\eta=0.3$ &  $-11.613$ & $-11.563$  & $0.42\%$  &     $-14.125$  & $-14.065$ & $0.42\%$ \\
		\hline 
	\end{tabular}
	\label{tab2}
\end{table}

\section{Numerical validation}\label{sec:num}

To validate the weakly nonlinear solutions, the membrane model is solved numerically using the Rayleigh--Ritz method \citep{ilanko2014rayleigh}. The main idea is to discretize the membrane energy functional \eqref{eq:mem} directly, rather than the associated equilibrium equations, and determine the unknowns from the stationarity condition of the resulting discrete energy.

Starting from the membrane energy functional \eqref{eq:mem}, we introduce the thickness stretch \(a\) as an independent variable to preserve the mixed variational structure and simplify the numerical implementation. The incompressibility constraint $
a\det(\nabla\bm r)=1$ is enforced through the Lagrange multiplier \(p\). The resulting mixed formulation involves the unknown fields \(r_1\), \(r_2\), \(a\), and \(p\).

Owing to the symmetry of the localized necking solution, only a quarter domain, 
	$[0,L_1/2]\times[0,L_2/2]$, 
	is considered in the numerical calculations. The symmetry conditions on the boundaries \(X_1=0\) and \(X_2=0\) are imposed as
	\begin{align}
	&r_1=0,\qquad \frac{\partial r_2}{\partial X_1}=0
	\qquad \text{on } X_1=0,\\
	&r_2=0,\qquad \frac{\partial r_1}{\partial X_2}=0
	\qquad \text{on } X_2=0,
	\end{align}
 The symmetry condition at \(X_1=0\) fixes the center of the localized solution and removes the translational degree of freedom. The remaining boundaries of the quarter domain are subjected to
	\begin{align}
	&\bm P\bm e_1=P\bm e_1
	\qquad \text{on } X_1=L_1/2,\\
	&\bm P\bm e_2=0
	\qquad \quad\, \text{on } X_2=L_2/2.
	\end{align}
 For the loading scenario with prescribed \(\lambda_1\), the corresponding tension \(P\) is determined from the relation $\eqref{eq:hom}_1$, and the same formulation is used with this equivalent loading condition.
	
The computational domain $[0,L_1/2]\times[0,L_2/2]$
is then discretized into \(n\times m\) square elements, where \(n\) and \(m\) denote the numbers of elements in the \(X_1\)- and \(X_2\)-directions, respectively, with mesh size $
h=L_1/(2n)=L_2/(2m)$. Within each element, the displacement components \(r_1\) and \(r_2\) are interpolated using bilinear shape functions, whereas the thickness stretch \(a\) and the Lagrange multiplier \(p\) are approximated by piecewise constant functions. Substituting these discrete approximations into the mixed variational formulation gives a discrete energy functional. The stationarity conditions of this discrete energy yield a nonlinear algebraic system for the nodal and element-wise variables.

The post-bifurcation solution is obtained by using the weakly nonlinear solution as an initial approximation for the nonlinear solver. The nonlinear algebraic system is solved in \textit{Mathematica} using the built-in \texttt{FindRoot} function, with convergence controlled by its internal accuracy and precision criteria. The mesh sensitivity of the numerical results is examined by successive mesh refinement, and no appreciable changes in the deformation profiles and loading parameters are observed upon further refinement. The results reported below are obtained using \(L_1=20\), \(L_2=1\), with the quarter domain discretized by a \(200\times 10\) mesh corresponding to a mesh size \(h=0.05\).

The necking solutions considered here are symmetric with respect to the mid-line \(X_2=0\). As shown in Figs.~\ref{fig:dp1} and \ref{fig:dp2}, the symmetric branch exists only in a small neighborhood of the bifurcation point. Away from the critical state, the symmetric and antisymmetric branches become increasingly close, making numerical continuation of the symmetric branch difficult over a large loading range. We therefore restrict attention to solutions close to the bifurcation point, which is also the regime in which the weakly nonlinear analysis is expected to be valid.

\begin{figure}[h!]
	\centering
	\subfloat[]{\includegraphics[width=0.435\textwidth]{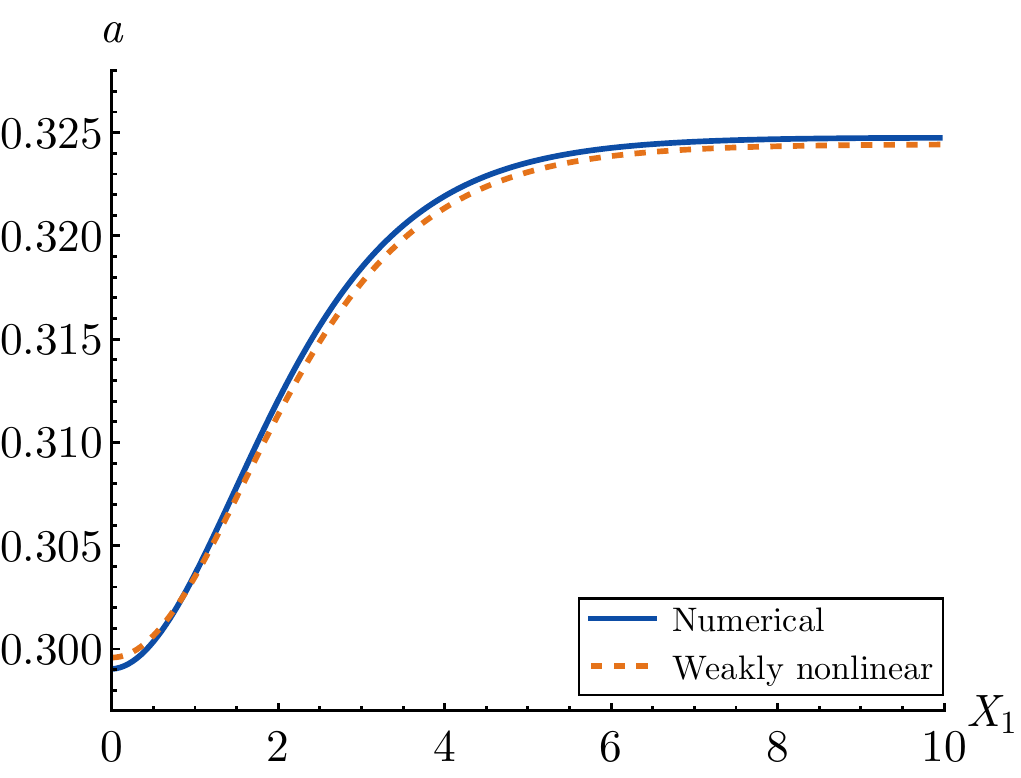}
	}\qquad\qquad
	\subfloat[]{\includegraphics[width=0.44\textwidth]{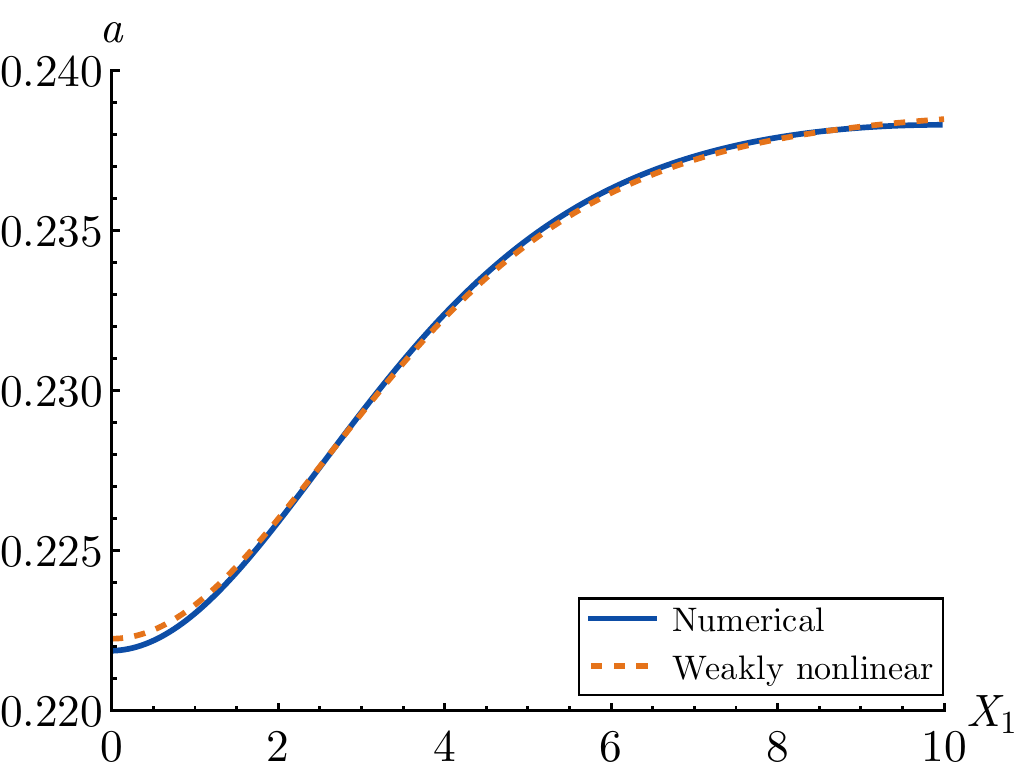}
	}
	\caption{Comparison between the numerical solution and the weakly nonlinear prediction for the thickness stretch \(a=z/Z\) under two loading scenarios. (a) Fixed tension \(P=2.5\), with the loading parameters \(\lambda_1=2.745\) (\(\lambda_{1\mathrm{cr}}=2.759\)), \(\lambda_2=1.123\), and \(U=0.3647\). (b) Fixed stretch \(\lambda_1=3.5\), with the loading parameters  \(\lambda_2=1.196\) (\(\lambda_{2\mathrm{cr}}=1.221\)), \(U=0.3047\), and \(P=3.661\).}
	\label{fig:cp}
\end{figure}

Figure~\ref{fig:cp} compares the numerical solution with the weakly nonlinear solution for two representative loading scenarios. In both cases, the weakly nonlinear solutions are in excellent agreement with the numerical results. These results provide strong evidence that the weakly nonlinear analysis accurately captures the localized necking behavior near the bifurcation point.

\section{Conclusion}\label{sec:con}

We have investigated the instabilities of dielectric membranes under combined mechanical and electrical loading, with the aim of identifying the bifurcation mode associated with the Hessian instability. Starting from a fully three-dimensional nonlinear electroelastic formulation, we derived asymptotically consistent membrane and plate models and used them to perform linear and weakly nonlinear analyses.

A central result of this work is that the Hessian instability can be interpreted as a zero-wavenumber bifurcation. Through linear bifurcation analysis, we showed that the Hessian stability criterion coincides exactly with the bifurcation condition at $k=0$, thereby establishing a direct connection between the loss of positive definiteness of the energy and bifurcation theory. The deformation mode associated with this long-wavelength bifurcation was then identified through a weakly nonlinear analysis, which shows that the bifurcation is subcritical and leads to localized necking, or equivalently, inhomogeneous thinning of the membrane.

The dispersion relation also reveals an additional feature: the existence of both symmetric and antisymmetric branches. The weakly nonlinear solution corresponds to the symmetric branch and accurately describes the deformation in a small neighborhood of the bifurcation point. However, the symmetric branch exists only in a narrow neighborhood of the bifurcation point. As the loading parameter moves farther from the critical state, the antisymmetric branch becomes dominant. This suggests that the post-bifurcation behavior may involve a transition from symmetric localized thinning to an antisymmetric deformation mode, a feature that cannot be captured by the local weakly nonlinear analysis alone.

The role of reduced models has also been clarified. For the plane-stress configuration considered here, the membrane model accurately captures both the onset of instability and the associated weakly nonlinear behavior, while finite-thickness effects enter only as small higher-order corrections. The localized thinning predicted by the present analysis may represent the initial stage of the two-phase deformation reported by \cite{huang2012electromechanical}. These results provide a bifurcation -based interpretation of the Hessian instability and establish a simple theoretical framework for analyzing electromechanical instabilities in dielectric membranes.

\section*{Acknowledgements}

This work was supported by the National Natural Science Foundation of China (Grant Nos 12402068, 12472067), Guangdong
Basic and Applied Basic Research Foundation (Grant No 2023A1515111141) and Youth S\&T Talent Support Programme of Guangdong Provincial
Association for Science and Technology (No SKXRC2025437).

\appendix

\section{Coefficient matrices in the linear bifurcation analysis}\label{app:coe}

The nonzero entries of the coefficient matrices $M$ and $N$ appearing in \eqref{eq:coeM} are given below:

\begin{align}
\begin{split}
&m_{21}=-k^2\,
\frac{
	\lambda_1^8 \lambda_2^4
	+(J_m+3)\lambda_1^6 \lambda_2^4
	+\lambda_1^2 \lambda_2^2(4J_m+12-3\lambda_2^2)+\lambda_1^4 \lambda_2^2\big(\lambda_2^6-(J_m+3)\lambda_2^4-9\big)-2}{\lambda_1^8 \lambda_2^4+\lambda_1^4 \lambda_2^2+
	\lambda_1^6 \lambda_2^4(\lambda_2^2-J_m-3)
},\\
&m_{24}=-k^2\,
\frac{
	3\lambda_1^6\lambda_2^6
	+\lambda_1^4\lambda_2^2\big(\lambda_2^6-(J_m+3)\lambda_2^4-6\big)
	+\lambda_1^2\lambda_2^2(4J_m-5\lambda_2^2+12)
	-2
}{
	\lambda_1^7\lambda_2^5+\lambda_1^3\lambda_2^3 
	+\lambda_1^5\lambda_2^5(\lambda_2^2-J_m-3)
},\\
&m_{42}=\frac{
	2\lambda_2
	+2\lambda_1^2\lambda_2^3(3\lambda_1^2-2J_m-6)
	+5\lambda_1^2\lambda_2^5
	+\lambda_1^4\lambda_2^7(J_m+3-3\lambda_1^2)
	-\lambda_1^4\lambda_2^9
}{	2\lambda_1\big(
	\lambda_1^4\lambda_2^2\big(\lambda_2^6-2)+	2\lambda_1^2\lambda_2^2(J_m+3-2\lambda_2^2)-1\big)
},\\
&m_{43}=-k^2\,
\frac{
	\lambda_1^2\lambda_2^4\big(
	1+\lambda_1^4\lambda_2^2
	+\lambda_1^2\lambda_2^2(\lambda_2^2-J_m-3)
	\big)
}{2\lambda_1^4\lambda_2^2(\lambda_2^6-2)	+4\lambda_1^2\lambda_2^2(J_m+3-2\lambda_2^2)-2
},\\
&n_{12}=\frac{\mu J_m  \lambda_1^2 \lambda_2^2}{
	\lambda_1^2 \lambda_2^2 (J_m+3-\lambda_2^2)
	-\lambda_1^4 \lambda_2^2-1},\\
&n_{13}=k^2 \frac{\mu J_m\lambda_1\lambda_2^3}{\lambda_1^2\lambda_2^2(J_m+3-\lambda_2^2)-\lambda_1^4\lambda_2^2-1},\\
&n_{21}=-\frac{
	2\mu J_m\big(
	\lambda_2
	+\lambda_1^2\lambda_2^3(3\lambda_1^2-2J_m-6)
	+2\lambda_1^2\lambda_2^5
	+\lambda_1^4\lambda_2^7(J_m+3-2\lambda_1^2)
	-\lambda_1^4\lambda_2^9
	\big)
}{\lambda_1\big(
	\lambda_2
	+\lambda_1^2\lambda_2^5+\lambda_1^2\lambda_2^3(\lambda_1^2-J_m-3)
	\big)^2},\\
& n_{24}=\frac{
	2\mu J_m\big(
	-1
	+2\lambda_1^2\lambda_2^2 (J_m+3-2\lambda_2^2)
	+\lambda_1^4\lambda_2^2 (\lambda_2^6-2)
	\big)
}{\big(
	\lambda_2	+\lambda_1^2\lambda_2^5
	+\lambda_1^2\lambda_2^3(\lambda_1^2-J_m-3)
	\big)^2}.
\end{split}
\end{align}

\end{document}